\documentclass[smallextended]{svjour3}       

\smartqed  
 \usepackage{mathptmx}      
\usepackage{amsmath,amssymb,graphicx}
\usepackage{braket}
\usepackage{xspace}
\usepackage{graphicx}
\usepackage{caption}

\newcommand{\Yb}{\ensuremath{^{171}\rm{Yb}^{+}}\xspace}
\newcommand{\Ybi}[1]{\ensuremath{^{#1}\rm{Yb}^{+}}\xspace}
\newcommand{\fq}{\ensuremath{f_{\text{qubit}}}\xspace}
\newcommand{\fr}{\ensuremath{f_{\text{rep}}}\xspace}
\newcommand{\fv}{\ensuremath{f_{\text{vapor}}}\xspace}

\begin{document}

\title{Scalable Digital Hardware for a Trapped Ion Quantum Computer
\thanks{This work was supported by the Office of the Director of National Intelligence and Intelligence Advanced Research Projects Activity through the Army Research Office.}
}

\author{Emily Mount \and 
Daniel Gaultney \and 
Geert Vrijsen \and 
Michael Adams\and
So-Young Baek \and 
Kai Hudek \and
Louis Isabella \and
Stephen Crain \and
Andre van Rynbach \and
Peter Maunz\and
Jungsang Kim
}
\institute{J. Kim, E. Mount, D. Gaultney, G. Vrijsen, M. Adams, S-Y. Baek, K. Hudek, L. Isabella, S. Crain, A.~van~Rynbach \at
              Electrical and Computer Engineering Department, Duke University, Durham, N.C. 27708, USA\\
              Tel.: +1-919-660-5258\\
              Fax: +1-919-660-5247\\
              \email{jungsang@duke.edu} \\
           \and
           P. Maunz \at
              Sandia National Laboratories, Albuquerque, N.M. 87123
}

\date{Received: date / Accepted: date}
\maketitle

\begin{abstract}
Many of the challenges of scaling quantum computer hardware lie at the interface between the qubits and the classical control signals used to manipulate them. Modular ion trap quantum computer architectures address scalability by constructing individual quantum processors interconnected via a network of quantum communication channels. Successful operation of such quantum hardware requires a fully programmable classical control system capable of frequency stabilizing the continuous wave lasers necessary for trapping and cooling the ion qubits, stabilizing the optical frequency combs used to drive logic gate operations on the ion qubits, providing a large number of analog voltage sources to drive the trap electrodes, and a scheme for maintaining phase coherence among all the controllers that manipulate the qubits. In this work, we describe scalable solutions to these hardware development challenges.

\keywords{Quantum computation \and Qubits \and Trapped ions}
 \PACS{03.67.-a \and 03.67.Lx }
\end{abstract}

\section{Introduction}
\label{intro}

Trapped ions exhibit many favorable properties for use as qubits~\cite{Wineland1998,Blatt2008}, such as long coherence times~\cite{Langer2005}, near perfect state preparation and readout~\cite{Myerson2008,Noek2013}, and high fidelity qubit gate operations~\cite{Benhelm2008,Brown2011,Ballance2014}. The Modular Universal Scalable Ion-trap Quantum Computer (MUSIQC) platform provides a scalable architecture where registers of ion qubits can be trapped in scalable surface electrode ion traps~\cite{Merrill2011,Moehring2011}. Each quantum register must be capable of  performing arbitrary quantum logic operations among the qubits in the register, as well as supporting the optical interface needed to communicate qubits between quantum registers through a photonic network~\cite{Monroe2013,Monroe2014}. 

Any ion trap quantum computer, and specifically the MUSIQC architecture~\cite{Monroe2013,Monroe2014}, requires a wide range of classical hardware to drive the quantum operations needed for the computation.  Scaling up from a small laboratory experiment to a large-scale processor will require scalable classical resources in addition to qubits. Here we present scalable hardware to perform some of the classical tasks required by an ion trap quantum computer. These tasks include: continuous wave (CW) laser frequency locking, frequency stabilization of pulsed laser beams for quantum logic gates, and low-noise digital-to-analog converter (DAC) systems for trapping and shuttling ions. We designed each of these systems to be inexpensive, compact, and extensible such that they can grow or be replicated to accommodate a much larger experiment. The techniques are developed for \Yb  hyperfine qubit levels separated by approximately 12.6~GHz, where resonant laser beams at 369.5~nm and off-resonant Raman beams detuned 14--33~THz from the $^2\text{S}_{1/2}$ to $^2\text{P}_{1/2}$ levels are used to manipulate the qubits. Many of these techniques can be extended to hyperfine qubits in other ion species.

\section{Offset frequency lock for CW laser stabilization}
\label{sec:olock}

The narrow spectral linewidths ($\leq 10$MHz) that are typical of atomic qubits and neutral atom experiments require lasers with high frequency specificity and stability. Current state-of-the-art laser systems that rely on the control of operating current and temperature are unable to sufficiently stabilize the frequencies of commercial laser diodes to meet these needs, so active locking to another frequency reference is often used. Commercially available vapor cells allow for easy stabilization of lasers addressing neutral atom transitions, but such a simple and inexpensive solution is not available to lasers addressing ion transitions (by virtue of their ionic nature), and other methods must be employed.  

One approach to frequency locking an external cavity diode laser (ECDL) is to use a second laser, locked to a vapor cell at frequency \fv (e.g. 780~nm for Rb vapor), and an optical cavity. The optical cavity is designed to support both \fv and light at the desired ECDL frequency (e.g. 369.5~nm for \Yb ions). The optical cavity length is stabilized to \fv while the cavity error signal is used to provide feedback to the piezo-controlled grating inside the ECDL, effectively transferring the vapor cell frequency reference to the ECDL. A second approach, where the ECDL is stabilized to the transitions of an ion plasma (for example, in a hollow-cathode lamp) can also provide the frequency stability necessary for trapped ion experiments~\cite{Lee2014}. While both approaches can adequately stabilize the frequency of an ECDL for trapped ion experiments, scaling is a concern as they are expensive and cumbersome to implement, especially for a complicated experiment involving multiple lasers.  Here we present a scheme that transfers the lock of a single master laser to multiple slave lasers with minimal hardware, complexity, and cost.

\begin{figure*}
\includegraphics[width=0.5\columnwidth]{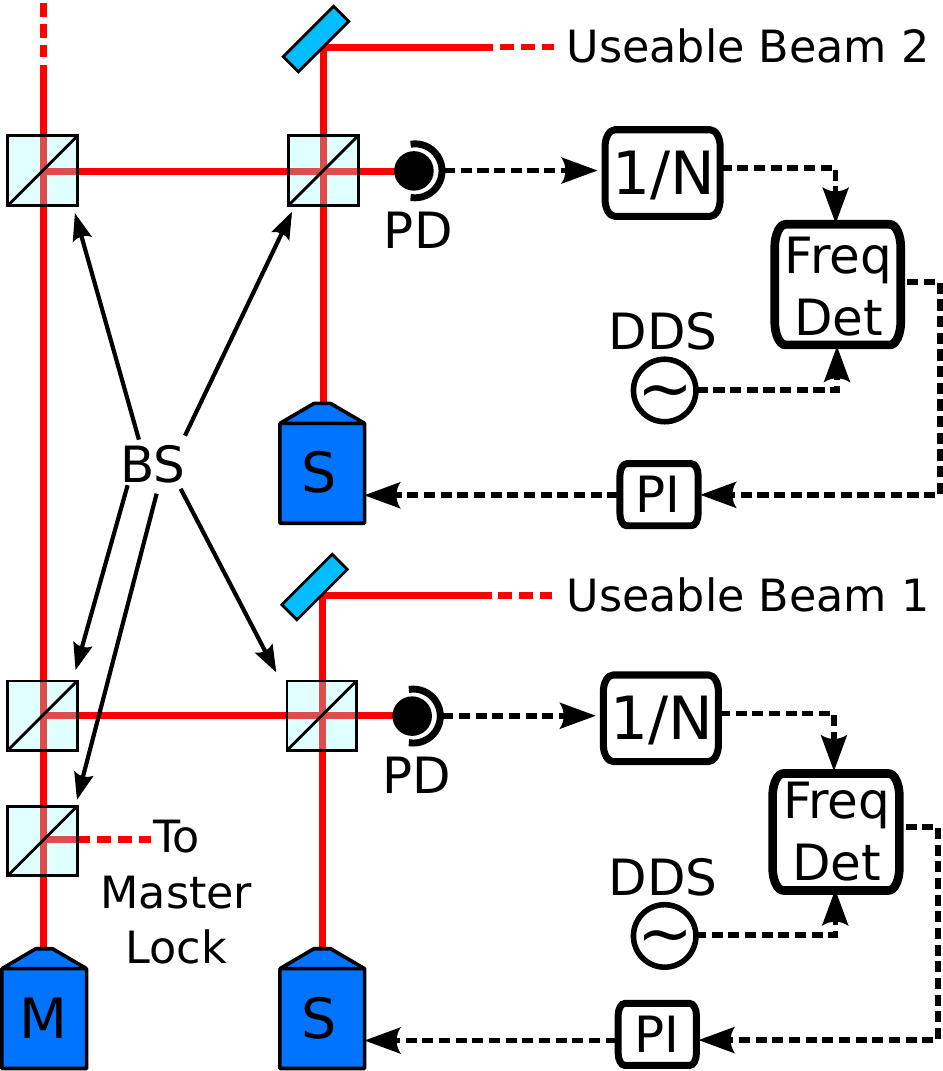}
\caption{Schematic of the offset lock setup.  A portion of the output power of an absolutely frequency stabilized master ECDL (M) is combined with a number of slave ECDLs (S) on a photodiode (PD).  The measured beat frequency is then divided by a $1/N$ prescaler ($N=10$, FPS-10-12, RF-Bay) and compared to the setpoint frequency output of a direct digital synthesizer (DDS) ($f_{\text{DDS}}$) by a frequency/phase detector.  The resulting error signal is filtered by a proportional-integral (PI) controller and fed back to the piezo-cotrolled grating of the slave ECDL to equalize the beat and setpoint frequencies making the frequency offset between the master and slave laser $N \times f_{\text{DDS}}$. }
\label{fig:g1}
\end{figure*}

In this scheme (shown in Fig.~\ref{fig:g1}), a portion of the slave's output and a small amount of light from the frequency stabilized master laser are matched in mode shape and polarization, and overlapped on a photodiode (PD). The overlapped electric fields result in a beat note at the difference frequency between the master and slave lasers, which is detected by PD. The range of detectable beat note frequencies is limited by the PD bandwidth and is typically a few tens of~MHz to several~GHz.  The beat note is amplified and compared to a reference signal (generated by a direct digital synthesizer (DDS) in our case) using a  phase/frequency detection circuit.  After appropriate filtering, the output of the circuit can be used as an error signal to provide feedback to the piezo-tunable grating and/or current of the slave ECDL using standard Pound-Drever-Hall control techniques~\cite{Drever1983}. The frequency difference between the master and slave lasers is digitally programmable by the DDS frequency to be $N \times f_{\text{DDS}}$ using a $1/N$ prescaler, as shown in Fig.~\ref{fig:g1}.

\begin{figure*}
\includegraphics[width=.7\columnwidth]{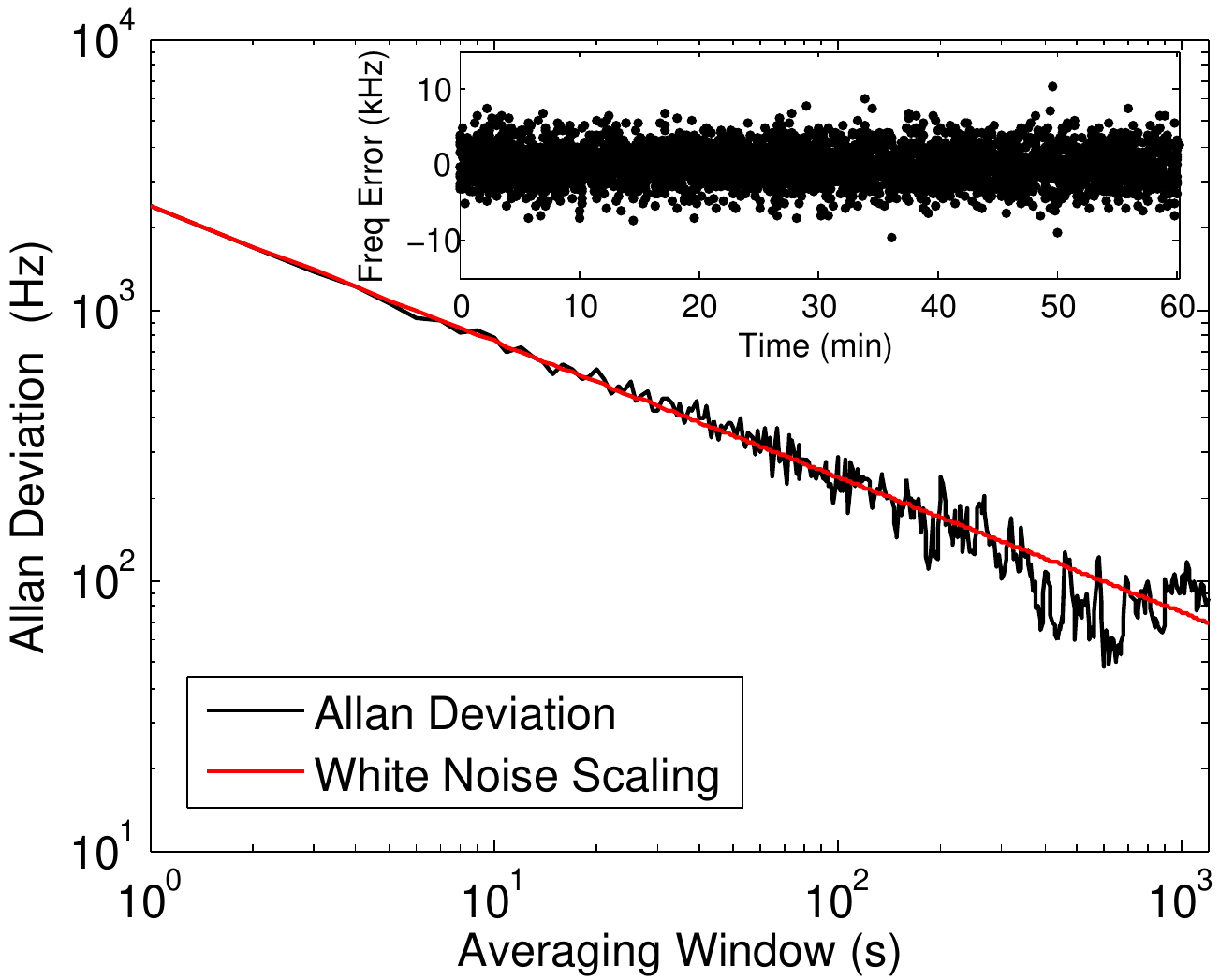}
\caption{Allan deviation of the beat note frequency as a function of averaging window time for a stabilized 369.5~nm ECDL.  The Allan deviation shows $1/\sqrt{t}$ scaling and is indicative of white frequency noise.  Inset: The difference between the locked beat note frequency and target frequency as a function of time for a 1 second averaging window as measured by a frequency counter.  Data shown is one data point per second for one hour.}
\label{fig:g2}
\end{figure*}

This solution is highly scalable as many slave lasers can be locked to a single master laser.  The number of slave lasers that can be locked to a single master laser is limited by the total power of the master laser. In our system, a 369.5~nm ECDL, with an output power of 3.4 mW, serves as the master laser which is frequency stabilized to a hollow-cathode lamp~\cite{Lee2014}. From the master laser, 600 $\mu$W of power is used to  lock 3 slave lasers at different offset frequencies (determined by the DDS frequency). We are currently capable of locking up to 8 slave lasers to a single master laser, which is more than enough to meet our current experimental demands.  If necessary, this number could be pushed higher with further improvements in mode-matching, or with more specialized equipment. The beat note is detected by a high-speed response (2~GHz)  photodiode (S9055, Hamamatsu), thus we are able to stabilize the master laser to the \Ybi{176} and have slave lasers stabilized to either the \Yb and \Ybi{174} resonances as necessary. From the Allan deviation of the beat note frequency for a slave 369.5~nm ECDL, we see that the beat note frequency noise is approximately white (Fig.~\ref{fig:g2}).

The footprint and complexity of the lock is reduced significantly using this approach as each additional slave laser merely requires a beam splitter to overlap the two laser beams, a focusing lens, and a photodiode, all of which can easily fit on a small breadboard with beams delivered via optical fibers. Additionally, the cost to implement the offset lock is more than an order of magnitude smaller than the cost of alternative approaches such as an additional transfer cavity or a hollow-cathode lamp.

\section{Digital optical frequency comb lock for Raman transitions}
\label{sec:bnlock}

For trapped ion qubits, driving qubit transitions using optical fields is convenient as they can be tightly focused for the individual addressing of single ions in a closely packed linear chain of ions. Optical frequency combs from mode-locked lasers are ideal for this purpose due to their broad bandwidth (10~GHz to 100~THz) which can bridge the wide energy separation of atomic states, and tightly spaced comb teeth, which are separated by the laser repetition rate (\fr), usually in the 50--1,000~MHz range~\cite{Hayes2010,Islam2014}.

\begin{figure*}
\includegraphics[width=0.99\textwidth]{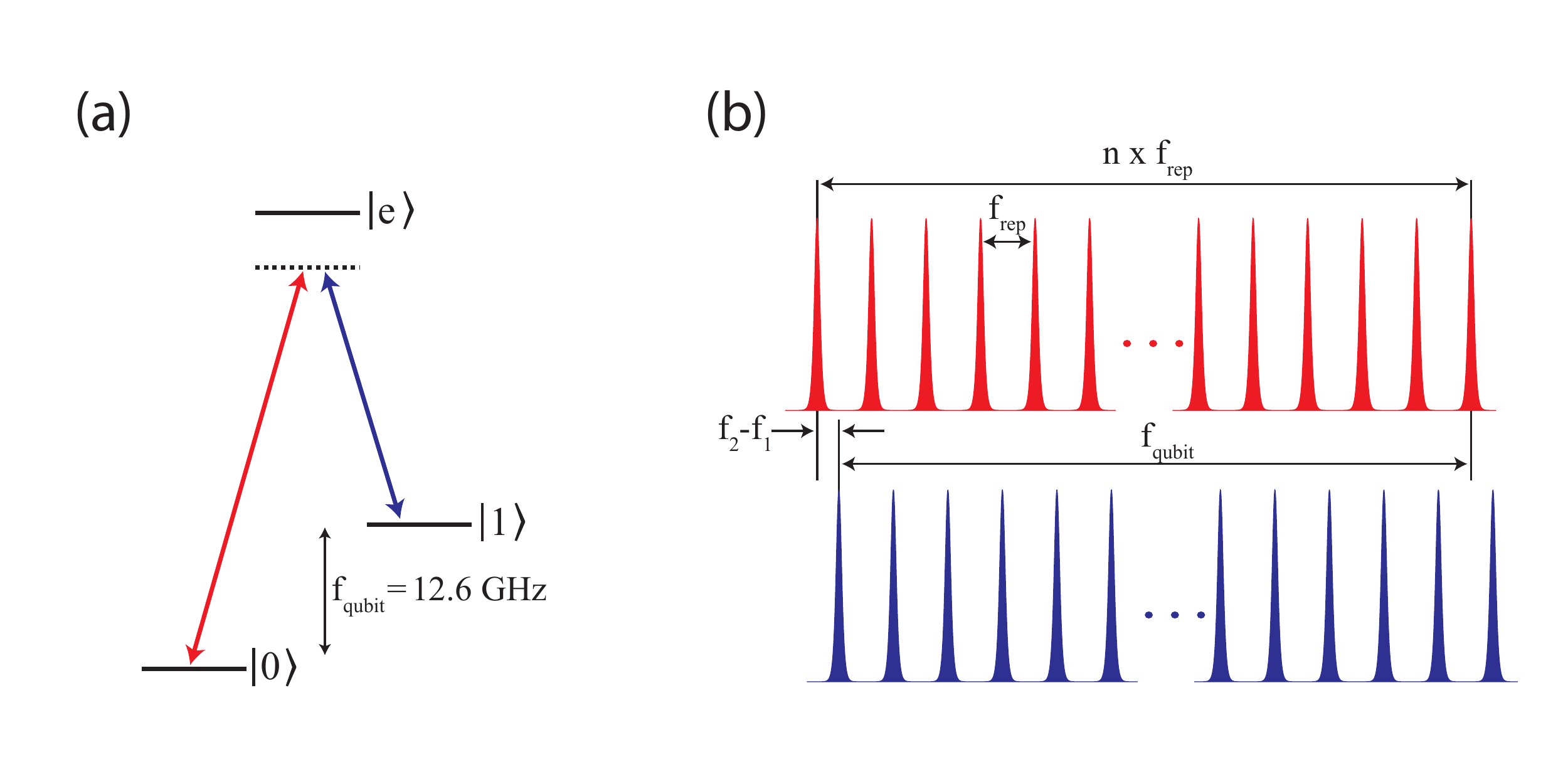}
\caption{(a) Energy level diagram of a three level $\Lambda$ system. Here, two of the hyperfine ground states of \Yb, separated by 12.6~GHz, are used as the qubit states $\ket{0}$ and $\ket{1}$. (b) Optical frequency combs from a single mode-locked laser. The combs are shifted relative to each other to ensure that the ion is driven by a pair of comb teeth with a frequency difference equal to the qubit frequency \fq.}
\label{fig:e1} 
\end{figure*}

The magnetically insensitive hyperfine ground states of the \Yb ion serve as our qubit states with $\ket{0}\equiv ^2$S$_{1/2}{\left|F=0, m_f=0\right>}$  and $\ket{1}\equiv^2$S$_{1/2}{\left|F=1,m_f=0\right>}$, at a qubit frequency \fq of 12.6~GHz. This system can be approximated as a three level $\Lambda$ system, shown in Fig.~\ref{fig:e1}(a). Two-photon stimulated Raman transitions between qubit states $\ket{0}$ and $\ket{1}$ can be driven using two optical frequency combs~\cite{Hayes2010}. Fig.~\ref{fig:e1}(b) shows optical frequency combs generated from a single mode-locked laser where the relative offset between the two combs, $f_2 - f_1$, is introduced by an acousto-optic modulator (AOM).  When the frequency difference $f_1 - f_2$ is such that
\begin{equation}
\fq=n \times \fr \pm (f_1 - f_2)
\label{eqn:1}
\end{equation}
where $n$ is an integer, an off-resonant two-photon Raman process induces the transition between $\ket{0}$ and $\ket{1}$ states through the virtual excited state $\ket{e}$. Because the Raman transitions are driven by the comb teeth pairs separated by exactly \fq,  it is sufficient to stabilize the relevant pair of comb teeth to each other, instead of stabilizing the absolute frequency or repetition rate of the laser output.

Thermal drifts and the vibration of mechanical parts in the laser cavity are among the main causes of drift in the laser repetition rate, which decreases gate fidelities by detuning the frequency comb pairs from \fq. Stabilization of \fr can be accomplished by directly stabilizing the laser cavity length; While this method provides sufficient stability and was used in previous experiments~\cite{Mount2013}, the lock bandwidth is typically limited to the~kHz range due to its reliance on piezo-translation of a relatively large mirror~\cite{HayesThesis}. This method requires access to the laser cavity, not available in many commercial lasers used in these experiments, and requires a stable microwave oscillator that filters near \fq which can be costly. Our alternative approach is to measure the drift of \fr by locking a digitally-controlled oscillator (such as a DDS) to the repetition rate signal using a digital phase-locked loop (PLL), and compute the driving frequency, $f_2$, of the optical modulator to satisfy Eq.~\ref{eqn:1}. We implement the lock digitally, using a proportional-integral (PI) controller implemented in a field programmable gate array (FPGA). The system provides a low-cost and scalable alternative to hardware-intensive analog methods and is a viable alternative when laser cavity access is limited.

\subsection{Implementation}
\label{sec:pi}
\begin{figure*}
\includegraphics[width=0.99\textwidth]{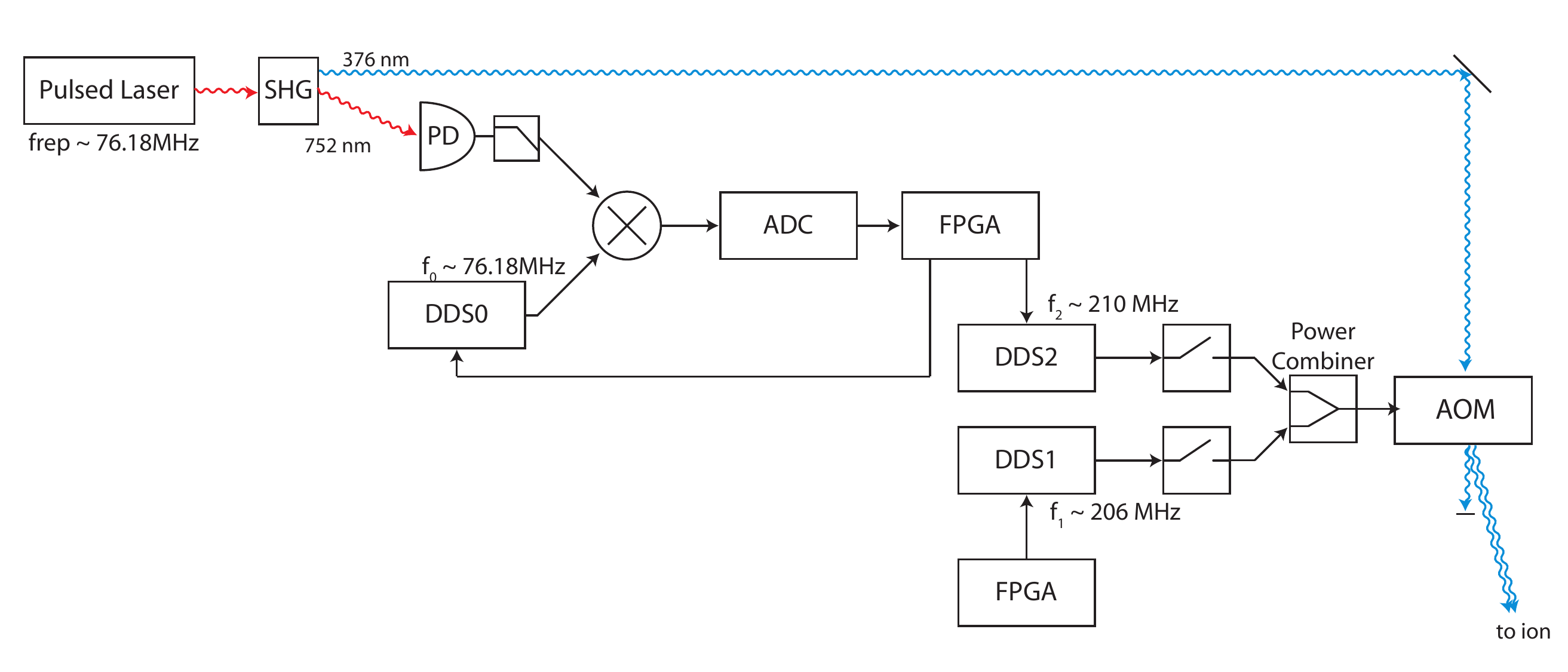}
\caption{Schematic of setup used to stabilize the frequency difference between the pairs of  comb teeth that drive the Raman transition. Laser light is incident on a photodiode (PD), which is used, in conjunction with a mixer, to lock the frequency of DDS0 ($f_0$) to the repetition rate (\fr).  The FPGA sets the frequency of DDS2 ($f_2$) based on $f_0$ to keep the difference frequency of the relevant frequency combs, shifted by $f_1$ and $f_2$, constant at \fq. SHG: second harmonic generation, DDS: direct digital synthesizer, ADC: analog-to-digital converter, PD: photodiode, FPGA: field programmable gate array, AOM: acousto-optic modulator.}
\label{fig:e2} 
\end{figure*}

The optical frequency comb stabilization setup is shown in Fig.~\ref{fig:e2}. A mode-locked titanium sapphire laser (centered at 752~nm, 2 ps FWHM pulse duration) is used to generate an optical frequency comb with comb teeth separation of $\fr \approx$ 76~MHz. The laser output is frequency doubled via second harmonic generation (SHG) using a BiBO crystal (length 8~mm). The light from the SHG output (376~nm) is split into two approximately co-propagating frequency combs by driving an AOM with two frequencies: $f_1$, controlled by the main experimental control FPGA and used to scan the detuning of the Raman beams in the experiments, and $f_2$, used to stabilize the frequency difference between the pairs of  comb teeth that drive the Raman transition. The co-propagating beam geometry ensures temporal overlap of the pulses at the ion.

A small fraction of the laser output at the fundamental wavelength (752~nm) is incident on a photodiode (PD, S9055 Hamamatsu), and then amplified and  low-pass filtered to attenuate the higher harmonics of \fr. The filtered PD output is mixed with the amplified output of a DDS (DDS0, Analog Devices, AD9912, 4 $\mu$Hz frequency resolution). The mixer output is low-pass filtered to eliminate the high frequency harmonics, while preserving the portion related to the frequency and phase difference, which provides the error signal for the PI controller in the PLL. Digitization of the error signal is performed by an analog-to-digital converter (ADC, Analog Devices AD7671, 16-bit resolution, 1 MSPS) and then read by the FPGA (Altera Cyclone II). The PI computation is performed in the FPGA which sets the frequency values of DDS0 ($f_0$) to track \fr, and DDS2 ($f_2$) to satisfy Eq.~\ref{eqn:1}. All DDSs share the same reference clock.

The lock operates as follows: a small change in \fr, in time step k, results in a digitized error signal  $e_k \propto (\fr - f_0) \Delta t+(\phi_{rep}-\phi_0)$, where $\Delta t$ is the sampling interval, and $\phi_{rep}$ and $\phi_0$ are the phases of the PD and DDS0 outputs, respectively. Given constants $P$ and $I$, each time step a correction is made to $f_0$ where 
\begin{equation}
f_0(k+1)=f_0(k)+(P \times e_k) +( I \times \sum_{n=1}^{k} e_n).
\label{eqn:2}
\end{equation}
This locks the frequency and phase of $f_0(t)$ to $\fr(t)$ via feedback control. Consequently, when \fr drifts such that
\begin{equation}
\fr(k+1)=\fr(k)+\delta,
\label{eqn:3}
\end{equation}
$\delta$ is known, as the corresponding $f_0$ is set by the FPGA to match \fr. The FPGA feeds-forward the value of the $n^{th}$ harmonic of $\delta$ to $f_2$, so that
\begin{equation}
f_2(k+1)=f_2(k)+n \times \delta.
\label{eqn:4}
\end{equation}
This maintains the equality in Eq.~\ref{eqn:1}, and
\begin{equation}
\fq = n \times[ \fr(k+1)+ \delta] + (f_1 - [f_2(k)+n \times \delta]),
\label{eqn:5}
\end{equation}
allowing for stable qubit transitions as \fr drifts. With $\fr \approx 76$~MHz and $\fq \approx 12.6$~GHz, $n = 166$.

\subsection{Lock parameters}

\begin{figure*}
\includegraphics[width=\textwidth]{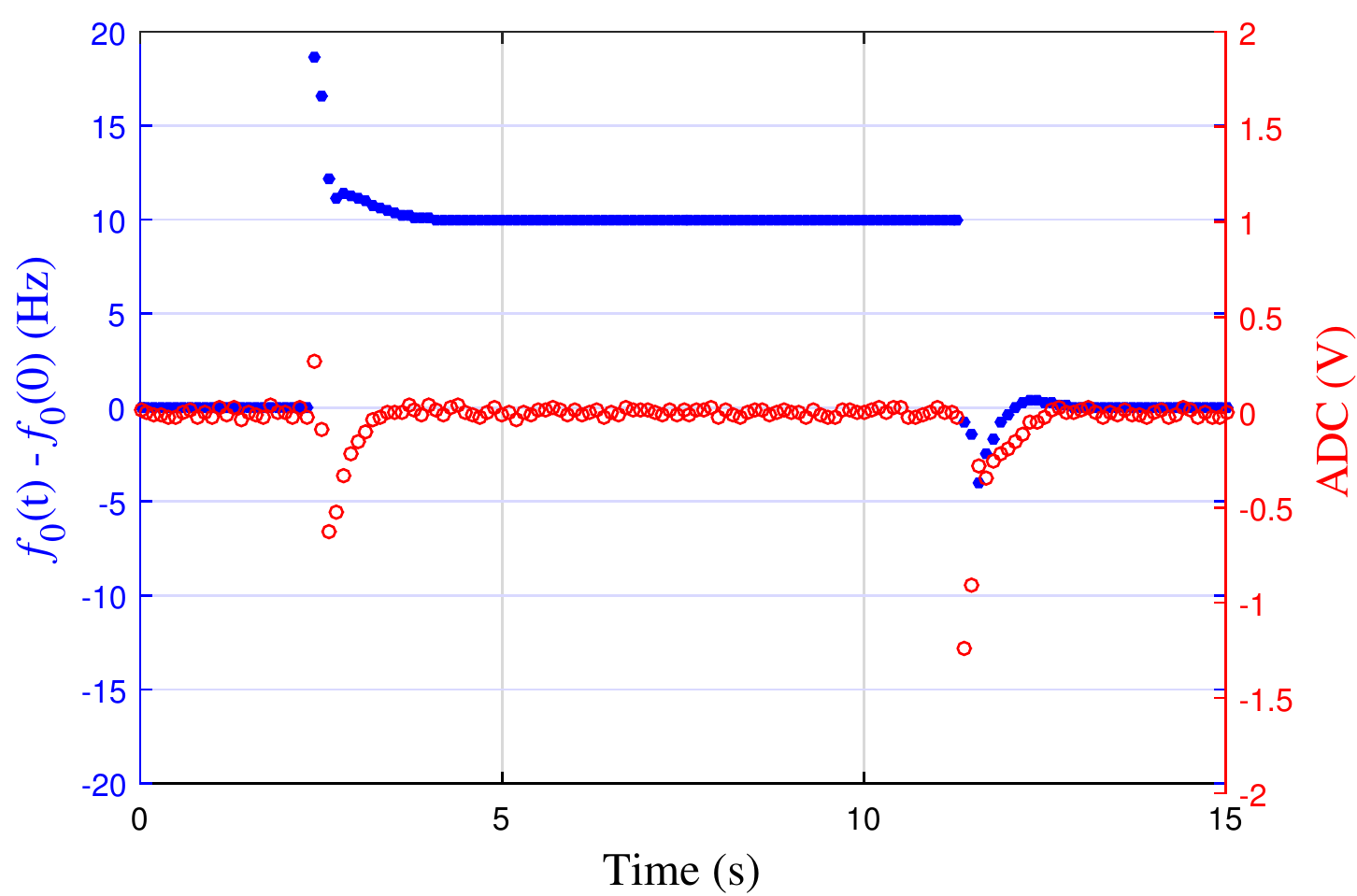}
\caption{Lock response to a finite \fr step. By replacing the PD in Fig.~\ref{fig:e2} with a DDS, we can measure the reaction of the lock to a finite jump in \fr for the chosen $P$ and $I$ coefficients. We choose an $I$ that, with $P=1$, will produce a slew rate of $\sim 50$~Hz/s. The difference between the current and initial DDS frequency is shown as filled blue circles, and the open red circles represent the ADC value read.}
\label{fig:e3} 
\end{figure*}

By examining the mixer output (error signal) on an oscilloscope, we determined a maximum rate of \fr drift of $\approx 5$~Hz/s. To establish the optimal integral gain coefficient, $I$, and the frequency difference detection sensitivity of our system, we replaced the PD output with another DDS, which was amplified to the same output power level and set to about the same frequency as the PD output. After enabling the lock with a proportional coefficient $P=1$, the DDS frequency was changed abruptly, as shown in Fig.~\ref{fig:e3}. The coefficient $I$ was chosen to provide a 50 Hz/s slew rate, sufficient for the lock to follow our expected 5 Hz/s rate of change. With the $I$ determined, to minimize noise we used the minimum proportional coefficient $P$ necessary to stabilize the lock. 

\subsection{Digital filtering}
\label{sec:digitalFiltering}

\begin{figure*}
\includegraphics[width=\textwidth]{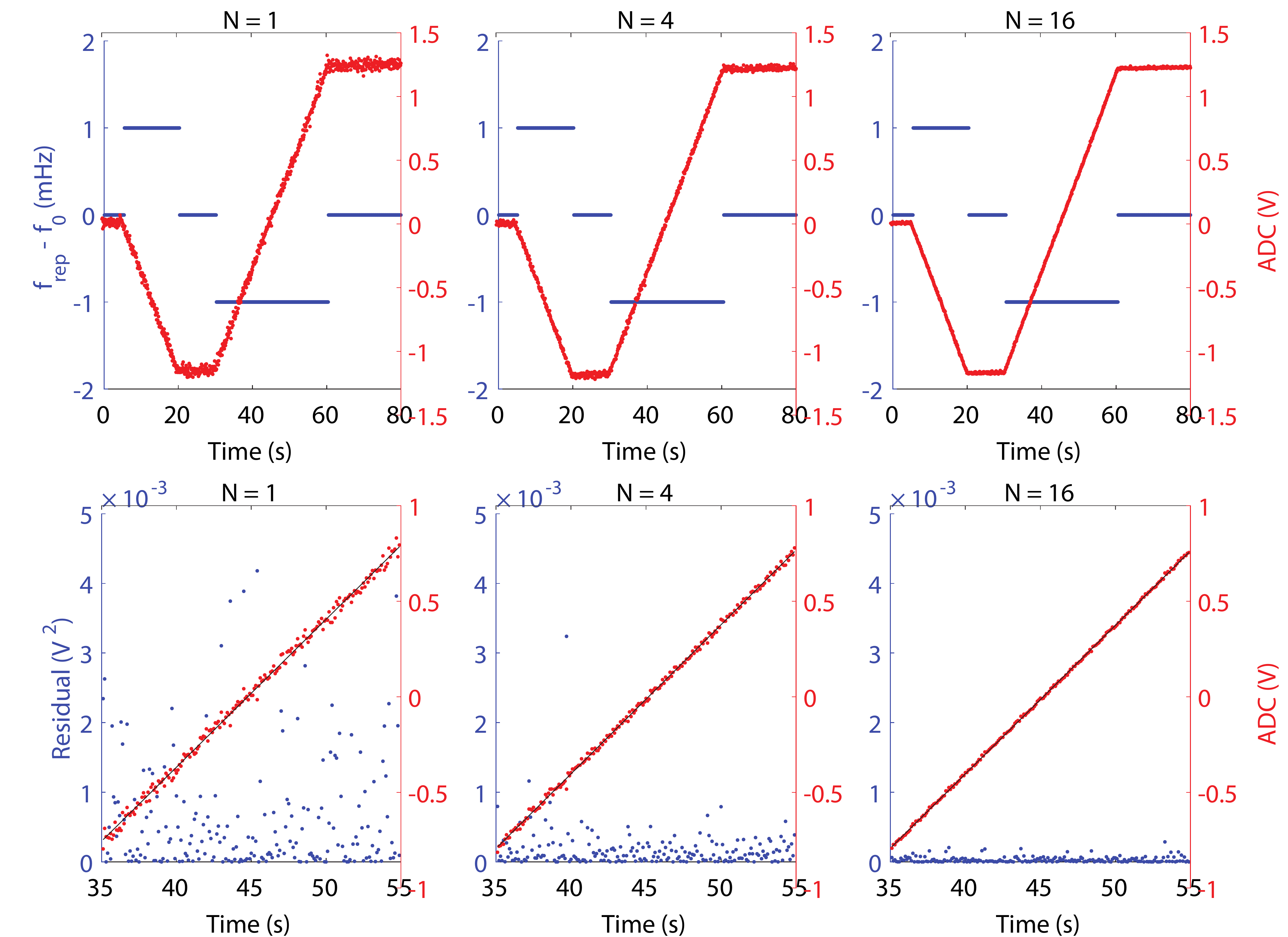}
\caption{Digital averaging. With the frequency comb stabilization system unlocked the PD is replaced with a DDS, set to take finite frequency steps. (top row) When $\fr-f_2 \ne 0$ the error deviates from zero until $\fr-f_2 = 0$ again. (bottom row) The error signal noise is reduced by digital averaging. Using a simple linear fit, we see R$^2$ values of 0.121, 0.034, 0.007 (V$^2$) for $N=$1, 4, and 16 respectively.}
\label{fig:e4} 
\end{figure*}

In our lock implementation, digital filters can be inserted without any hardware modification. A simple averaging filter was implemented where, upon receiving $N$ error signal samples at the maximum rate of 1 MSPS from the ADC, the PI loop proceeds using $\bar{e}_k=\frac{1}{N}\sum_{n=1}^{N} e_n$ as the error signal. Fig.~\ref{fig:e4}(a) shows the ADC readings where \fr is again replaced with a DDS. The DDS is set to take finite frequency steps, similar to Fig.~\ref{fig:e3} with the system unlocked, at different values of $N$. The ADC noise is clearly reduced by digital averaging as evidenced by the residuals of a simple linear regression, shown in Fig.~\ref{fig:e4}(b). While averaging decreases the error signal noise, the locking bandwidth is also reduced. With no digital averaging, the locking bandwidth of the digital PI loop is limited to 1~MHz by the conversion speed of the ADC. With digital averaging, the lock bandwidth is reduced by $1/N$.

\subsection{Lock verification}
High fidelity qubit operations require that Raman beat notes stay coherent with the qubit frequency for the duration of the computation. To measure the coherence time, a Ramsey experiment was performed. First, the qubit is initialized into the $\ket{0}$ state, followed by a $\pi/2$  pulse that rotates the qubit into a superposition state.  After a wait time of $\tau$, a second $\pi/2$ pulse, with a phase shift of $\phi$ with respect to the first pulse, is implemented followed by a measurement. This procedure is repeated for many values of $\phi$ resulting in a fringe. The fringe visibility indicates how well difference frequency of the Raman beams remains synchronized with the phase of the qubit.

\begin{figure*}
\includegraphics[width=\textwidth]{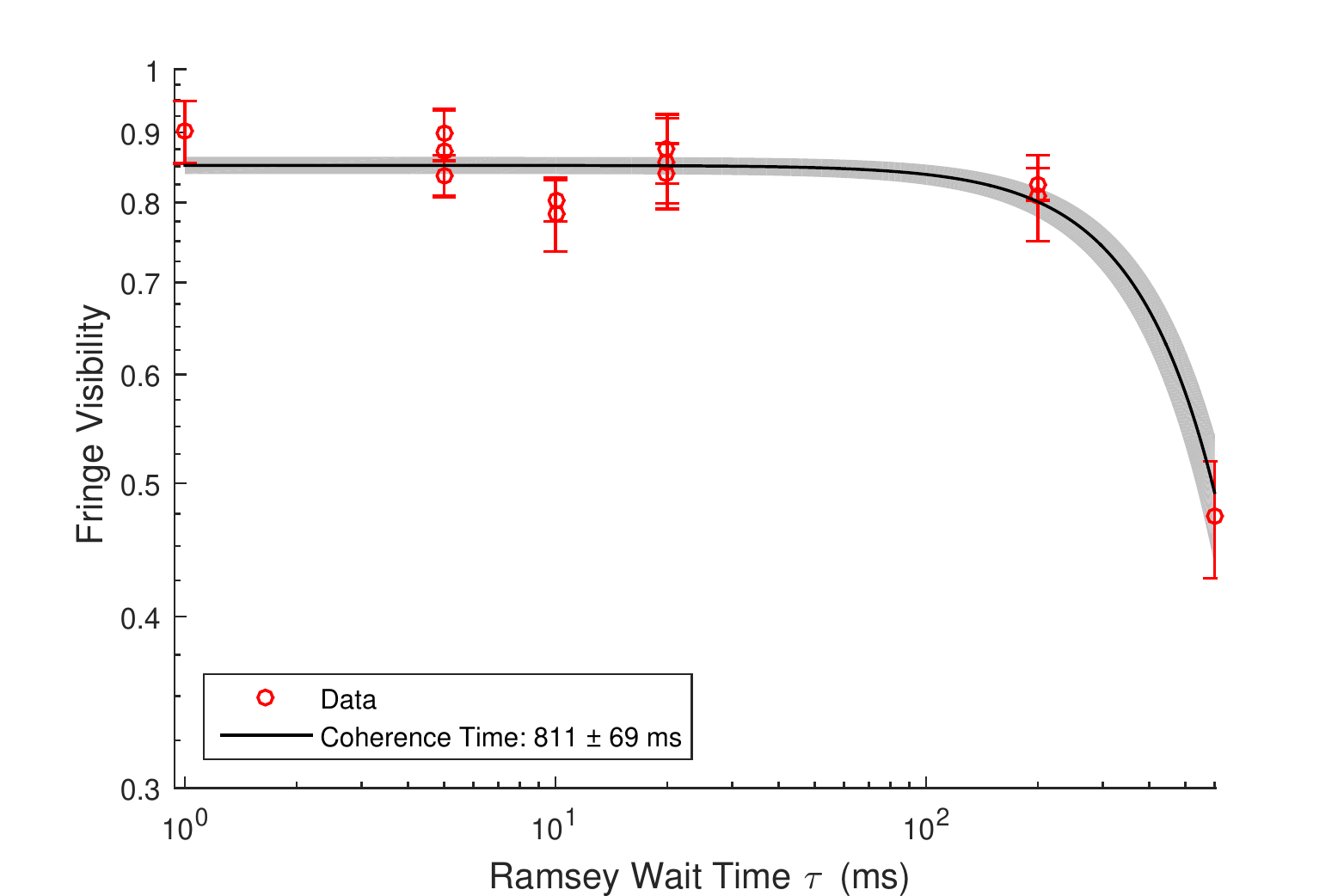}
\caption{Coherence time using the digital optical frequency comb lock. Red data points represent the Ramsey fringe visibility at various values of time delay, $\tau$. The fringe visibility as a function of $\tau$ is fit to a Gaussian function, black line, resulting in a coherence time of $811 \pm 69$ ms. Dark gray represents standard error on the fit.}
\label{fig:e45} 
\end{figure*}

Using the digital optical frequency comb lock we measure a coherence time of $811 \pm 69$ ms, shown in Fig.~\ref{fig:e45}. The coherence time is determined by fitting the Ramsey fringe visibility versus $\tau$ to the Gaussian function $A\times$exp(-$\tau^2/\alpha^2$), where $\alpha$ is the coherence time. By directly stabilizing the cavity length in the same laser, we previously measured a coherence time of $687 \pm 55$ ms using a similar procedure~\cite{Mount2013}.

\subsection{Global synchronization}

In a scalable quantum hardware system, it is crucial that the phase coherence of all classical fields that manipulate various parts of the system be maintained. In the case of Raman beams that drive quantum logic gates, the phase of the Raman beat notes must be maintained, in case there are multiple pairs of Raman beams driving quantum gates across the system. We ensure this by driving all the DDS controllers and microwave sources in the system from a single clock signal derived from an atomic standard. The relative phase difference between all of the RF and microwave control signals reaching the qubits must be calibrated (and stabilized) to operate a distributed quantum system such as MUSIQC.

\section{Digital optical intensity lock}
\label{sec:intlock}

\begin{figure*}
\includegraphics[width=\textwidth]{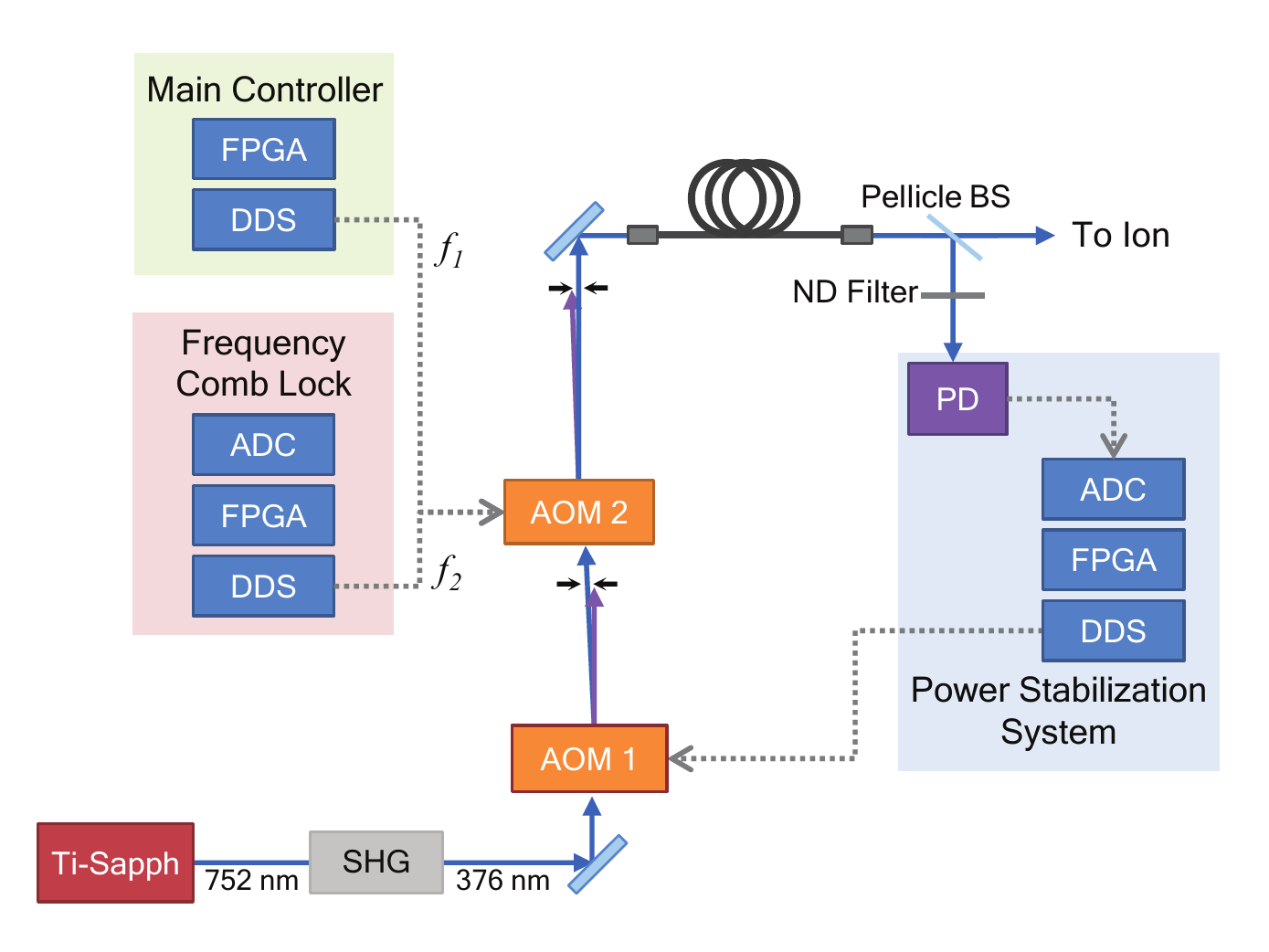}
\caption{Experimental setup with optical intensity lock. A small fraction of the Raman laser light is incident on a photodiode (PD), which provides the error signal for a PI controller implemented in an FPGA.   The FPGA dynamically sets the amplitude of the DDS signal driving AOM 1 to keep the intensity on the PD constant. }
\label{fig:e5} 
\end{figure*}

Intensity fluctuations of the Raman beams at the ion location result in the degradation of gate fidelities, and drifts in Rabi frequency and Stark shift calibrations. For high fidelity gates, control of the laser power at the ion location is critical. Here, we implement a power stabilization system using a PD, ADC, FPGA, and a DDS. As shown in Fig.~\ref{fig:e5}, after the frequency combs exit the fiber, a small amount of collimated light is picked-off and placed on a PD. The PD signal is digitized by an ADC (AD7671) and relayed to the FPGA (Altera, Cyclone II). This provides the error signal for a PI feedback loop, similar to that used in Sect.~\ref{sec:pi}. The FPGA actively stabilizes the power on the PD by controlling the amplitude of the DDS signal driving AOM 1. This implementation allows for gating of the PI loop by a digital trigger pulse, sent from the FPGA in our main controller. In most experiments the Raman beams must be turned on and off. A digital trigger allows the PD to be placed after the AOMs, closer to the ion, providing better intensity stabilization. In our experiments, we turn the Raman beams and PI loop on during the Doppler cooling cycle, allowing the power in the Raman beams to be stabilized before the experiment. Once the cooling cycle is complete, the feedback control is turned off, and we hold the final value of the AOM amplitude setting from the lock throughout the remainder of the experiment until the next cooling cycle. Using this sample and hold approach, slow intensity fluctuations due to beam pointing instability before the fiber are corrected. 

\begin{figure*}
\includegraphics[width=0.99\textwidth]{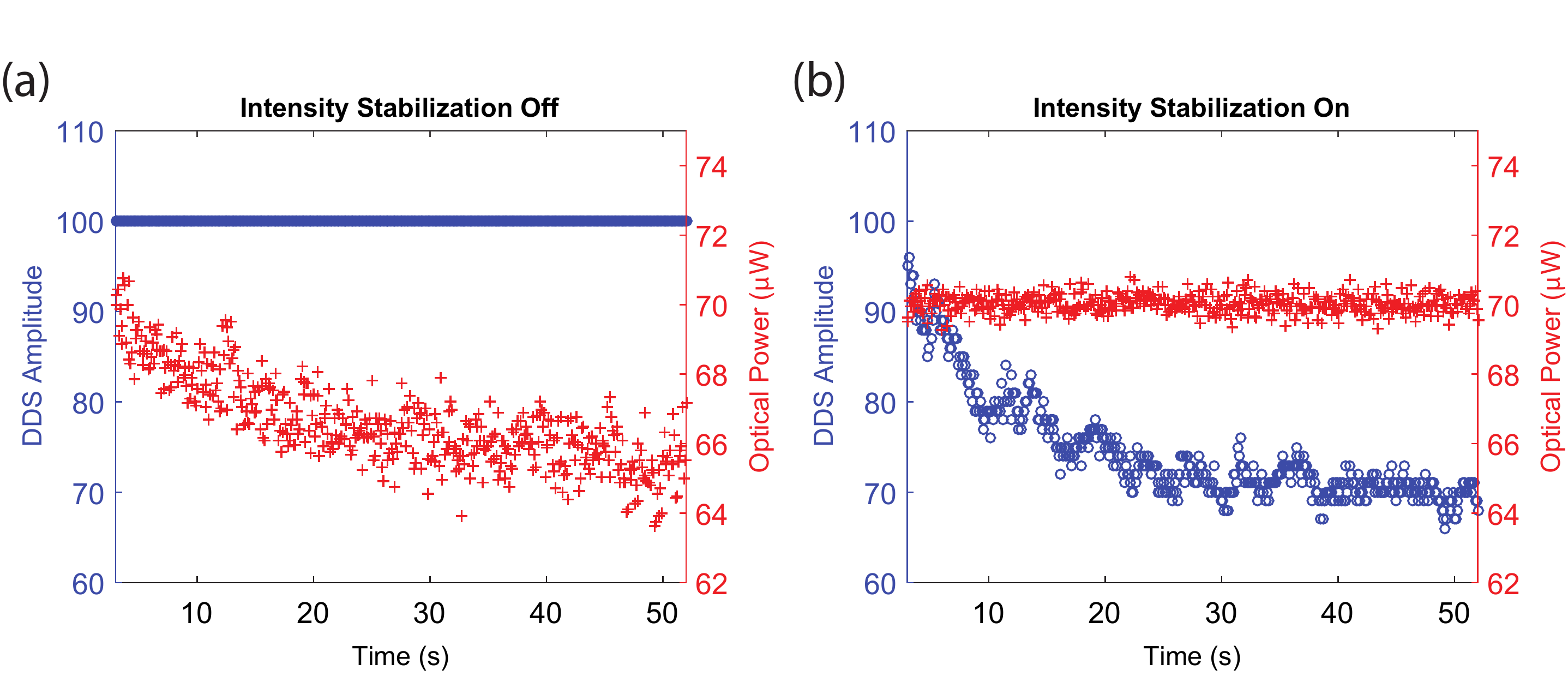}
\caption{Photodiode signal with and without digital optical intensity lock. The voltage read by the ADC (red crosses) is proportional to the optical intensity of the Raman beams. (a) When the intensity lock is off, the DDS amplitude (blue circles) is held constant and the ADC voltage, representing the optical power in the beam, drifts. (b) With the intensity lock on, the DDS amplitude changes to keep the ADC voltage constant.}
\label{fig:e6} 
\end{figure*}

Figure \ref{fig:e6} shows the impact of intensity stabilization. When the lock is turned off (constant DDS amplitude, as in Fig. \ref{fig:e6}a), the power delivered through the fiber drifts as monitored by the PD (sampled by the ADC). Once the locked is turned on, the DDS amplitude is adjusted to maintain constant optical power out of the fiber (Fig. \ref{fig:e6}b).

\section{Next generation lock}
\label{sec:nextgen}

We are developing a general-purpose 8-channel digital proportional-integral-differential (PID) control system to replace the PI loop implementations discussed in sections 2-4 and  to be used for future locking tasks. It supports multiple user-selectable outputs for the lock, such as DC voltage through digital-to-analog converters (DACs) and frequency or amplitude of DDS signals, among others.
 
\begin{figure*}
\includegraphics[width=.95\columnwidth]{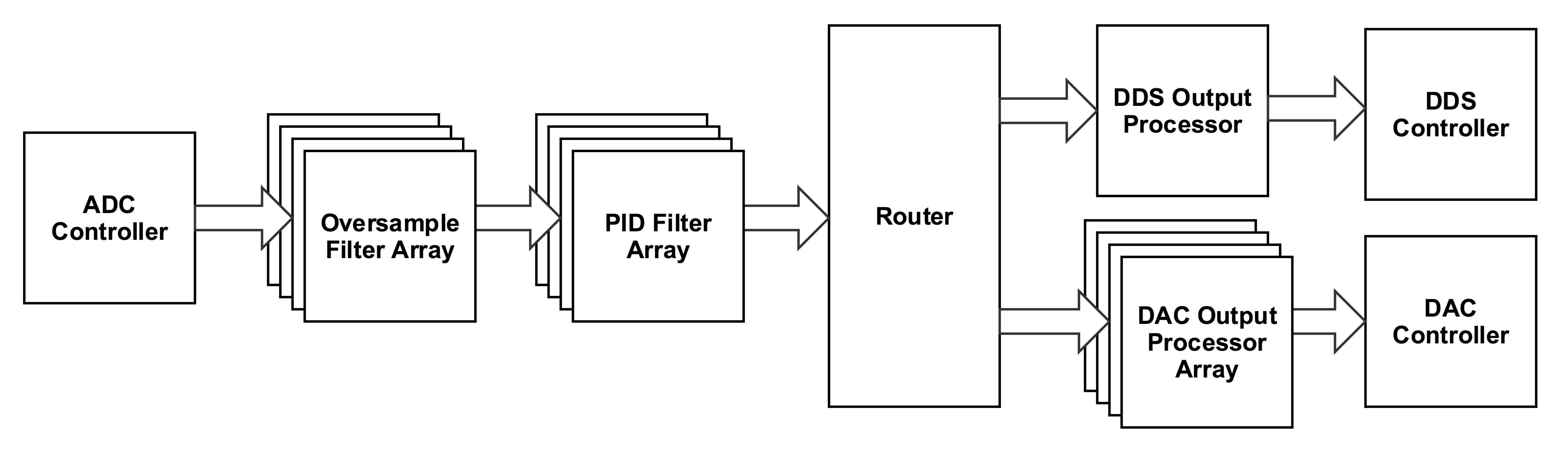}
\caption{Next generation PID system. The ADC controller reads eight channels of error data from the AD7608. The AD7608 runs at its maximum update rate of 200kHz. Error data is passed, one channel at a time, from the ADC controller into the PID processing pipeline. The first stage of the pipeline is an oversampling module. The module maintains a running average of data words as they are received. The average is passed to the PID filter stage when the channel oversample ratio is satisfied. The PID filter applies a discrete PID filter to the averaged error data and the result is routed to the appropriate output channels. The output processor enforces signal bounds and optionally applies a linear transformation before passing the final data signal to the appropriate DAC or DDS controller. }
\label{fig:m1}
\end{figure*}

The PID logic is implemented on a Xilinx Spartan 6 FPGA, detailed in Fig.~\ref{fig:m1}. The FPGA is mounted on a circuit board that also contains an 8-channel ADC (Analog Devices AD7608, 18-bit resolution, 200 ksps) and DACs (Texas Instruments DAC8568, 16-bit resolution) for each channel. The locking system can support eight PID locks concurrently, each with optional DC output, using the integrated DACs (66~kHz maximum update rate), or AC output, via an off-chip DDS (Analog Devices AD9912, 100~kHz maximum update rate). A Python graphical user interface (GUI) was designed for selection of locking parameters and real-time lock monitoring, with communication to the FPGA provided by a USB 2.0 link. Digital filtering, as in section~\ref{sec:digitalFiltering}, is available on each channel through an oversampling module that performs digital averaging of the error signal.

\section{Digital to Analog Converter System}
\label{sec:5}

Historically, macroscopic 4-rod RF-Paul ion traps have been used for trapped ion quantum information experiments~\cite{Turchette2000,Maunz2007,Benhelm2008}. However, in recent years surface-electrode ion traps have come into use, as they provide repeatable trap parameters and support increased complexity. The trapping potential  is constructed from a sum of  RF and DC components. In a surface-electrode ion trap, these fields are generated by applying voltages to microfabricated metal electrodes located on a planar surface at a distance on the order of $100~\mu$m from the ions.  The voltages applied to these electrodes can be modulated to shuttle the ions along the trap axis or split and re-combine ion chains with high precision~\cite{Blakestad2009,Bowler2012,Shu2014}.  While ion shuttling is central to many quantum computing schemes, this introduces the need for a multi-channel DC-voltage supply with precise timing control between channels.  Although only a single RF source is required for the RF trapping potential, typical surface traps require 40-100 DC voltages in order to accommodate the desired shuttling range and precise motional control of the ions. Any electric field noise near the motional frequencies of the trap (typically 1~MHz to 5~MHz) will cause unwanted heating of the ion motion in the trap. Existing commercial systems are expensive, cumbersome to program, and generate noise near the trap frequencies due to their synchronous update rate of $\sim 1$~MS/s~\cite{NIPXI6733}.

\begin{figure*}
\includegraphics[width=1\columnwidth]{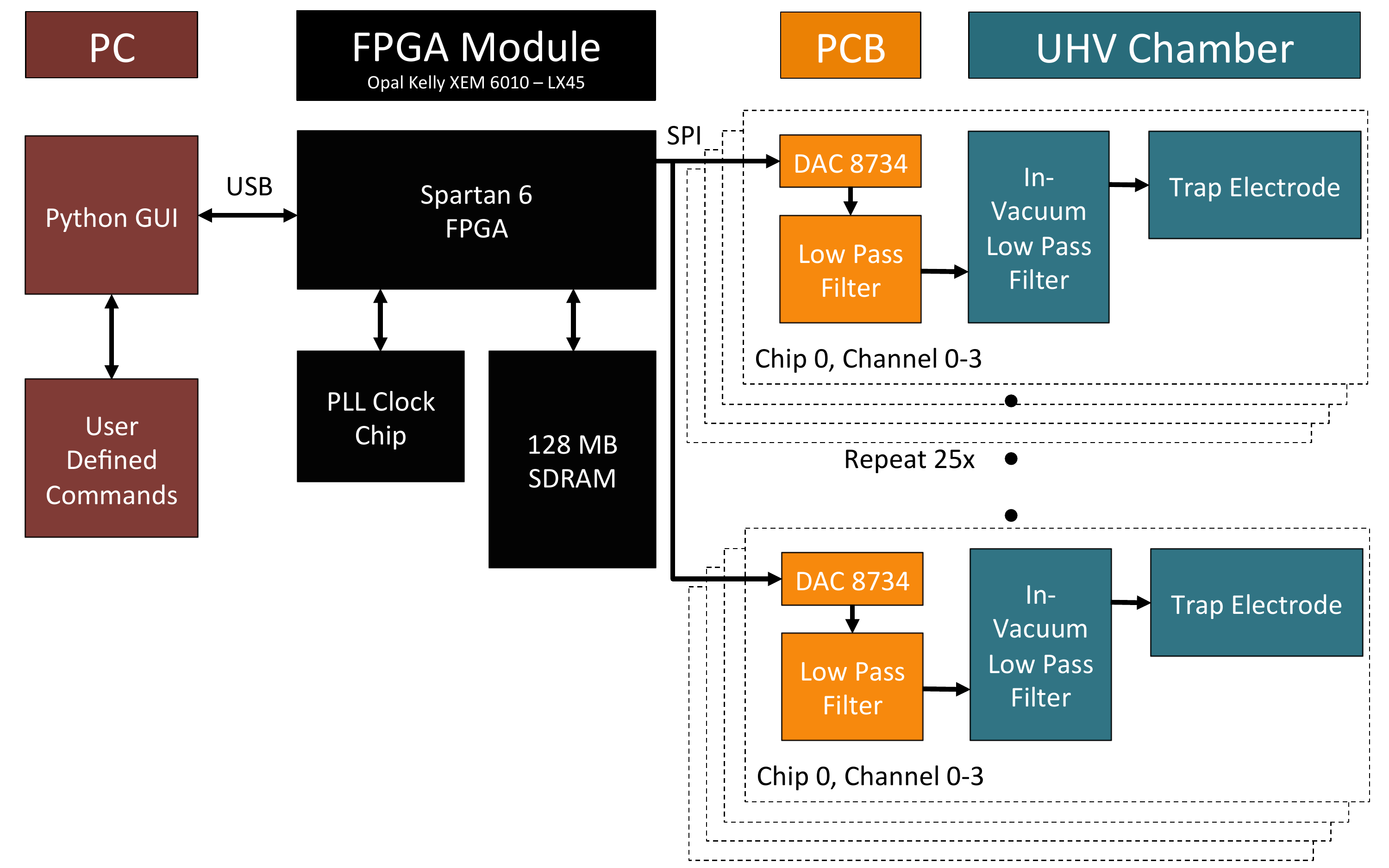}
\caption{A schematic overview of the 100-channel DAC system. The voltages are manipulated in real time by an Opal Kelly XEM6010 FPGA module which is controlled by a python GUI via a universal serial bus (USB) connection. The FPGA output pins are directly connected to 25 DAC8734 chips using a serial peripheral interface (SPI)-like protocol where each chip outputs 4 analog voltages. This SPI communication limits the update rate of the DACs to 430~kHz. The analog voltages go through several stages of RC filters before they arrive at the trap electrodes. These filters, which have a cutoff frequency of 150-400~kHz, reduce the noise on the trap electrodes near the resonant frequencies of the trap (around 1-5~MHz). The Opal Kelly module includes 128~MB of random access memory (RAM) which allows for multiple DAC output configurations with fast switching between configurations.  An aluminum heat-sink is attached to the DACs to dissipate up to 2~W of power per chip.}
\label{fig:DACSchematic}
\end{figure*}

In an effort to create a low-cost scalable voltage solution, we designed a DAC system with 25 DAC8734 chips, each with 4 channels (schematic shown in Fig~\ref{fig:DACSchematic}). An onboard FPGA module, the Opal Kelly XEM6010-LX45, communicates serially with each of these 25 DAC chips and enables real-time control of the voltages. This system achieves a five-fold cost reduction compared to a similar National Instruments DAC system, while also giving much better control over the update rates. Figure~\ref{fig:DACSchematic} also shows that this system can drive multiple 100-channel DAC systems, in case the overall quantum system involves multiple surface trap chips distributed over multiple vacuum chambers.

The DAC8734 chip was chosen primarily for its asynchronous nature. It has been observed that synchronous DACs whose update rates are near the ion trap frequencies cause ion heating~\cite{Blakestad2009} due to peaks in the electric field noise spectrum at the update rate and its harmonics. Most commonly used two-qubit gate schemes in ion trapping systems use the ions' shared motional state as a bus mode to generate entanglement between the ions.  This requires that the motional quanta of the bus mode stays constant throughout duration of the gate, and so the heating rate will significantly impact the achievable gate fidelity.  An asynchronous DAC has the advantage that updates can be paused while gates are being performed. In this way, the effect of the digital noise arising from the DACs on the fidelity of the two-qubit gates in the system can be minimized.  Alternatively, synchronous  DACs that update at a rate much faster than the ion trap frequencies have been found to cause little heating~\cite{Bowler2012}. However, these DACs are both expensive and require significant computational overhead, as FPGA clock frequencies are typically on the same order as the update rates. Asynchronous DACs can be updated at an arbitrarily low rate, well away from the ion trap frequencies. Other than ion shuttling, our DACs are typically not updated, effectively removing the noise associated with the DAC update rate during operations that are sensitive to motional heating. Table \ref{tab:DAC8734} shows some of the electrical properties of the DAC8734. The 16-bit resolution enables control over the voltage down to $\pm 300 \mu$~V, and the response time is a single clock cycle (20~ns). This response time is the time the DAC takes to respond to a latch command after an update has been sent (which takes 2000~ns, or 100 clock cycles).

\begin{table}
\caption{A table of the electrical characteristics of the DAC8934.}
\label{tab:DAC8734}
\begin{tabular}{| l | c |}
\hline
DAC Resolution &  16-bits \\ \hline
Max Output Voltage & $\pm 10$ V \\ \hline
Max Update Rate & 430 kHz  \\ \hline
Response Time & 20 ns \\ \hline
Digital Clock Frequency & $\leq 50$ MHz \\ \hline
Max Settling Time (Output changes from -10 V to 10 V) & 8 $\mu$s \\
\hline
\end{tabular}
\end{table}

For many ion trapping applications, it is important to be able to control the output of these DACs in real-time.  We make use of a commercial FPGA module (Opal Kelly XEM6010-LX45). Not only does this enable forward compatibility with future FPGAs, but it also simplifies the design process, as the FPGA module includes random access memory (RAM),  which enables memory-intensive operations such as ion shuttling. To control the system as a whole, we make use of multiple layers of software and hardware (shown in Figure \ref{fig:DACSchematic}). A Python program allows the user to generate a series of voltage sets to be executed. This program passes assembly code to the FPGA which is executed in real time and controls the serial outputs to the DAC chips. The voltage sets are sent from the PC to a  combination of block memory on the FPGA (32 sets maximum) and RAM (8 M sets maximum) on the Opal Kelly module. When the system is not in its shuttling mode, the FPGA uploads one set of voltages to the DAC chips and then turns off all outbound clocks to reduce the noise on the DAC outputs. The FPGA interpolates between these sets at a user defined update rate ($\leq 430$~kHz) and step size. 

\subsection{DAC noise analysis}
\begin{figure*}
\includegraphics[width=1\columnwidth]{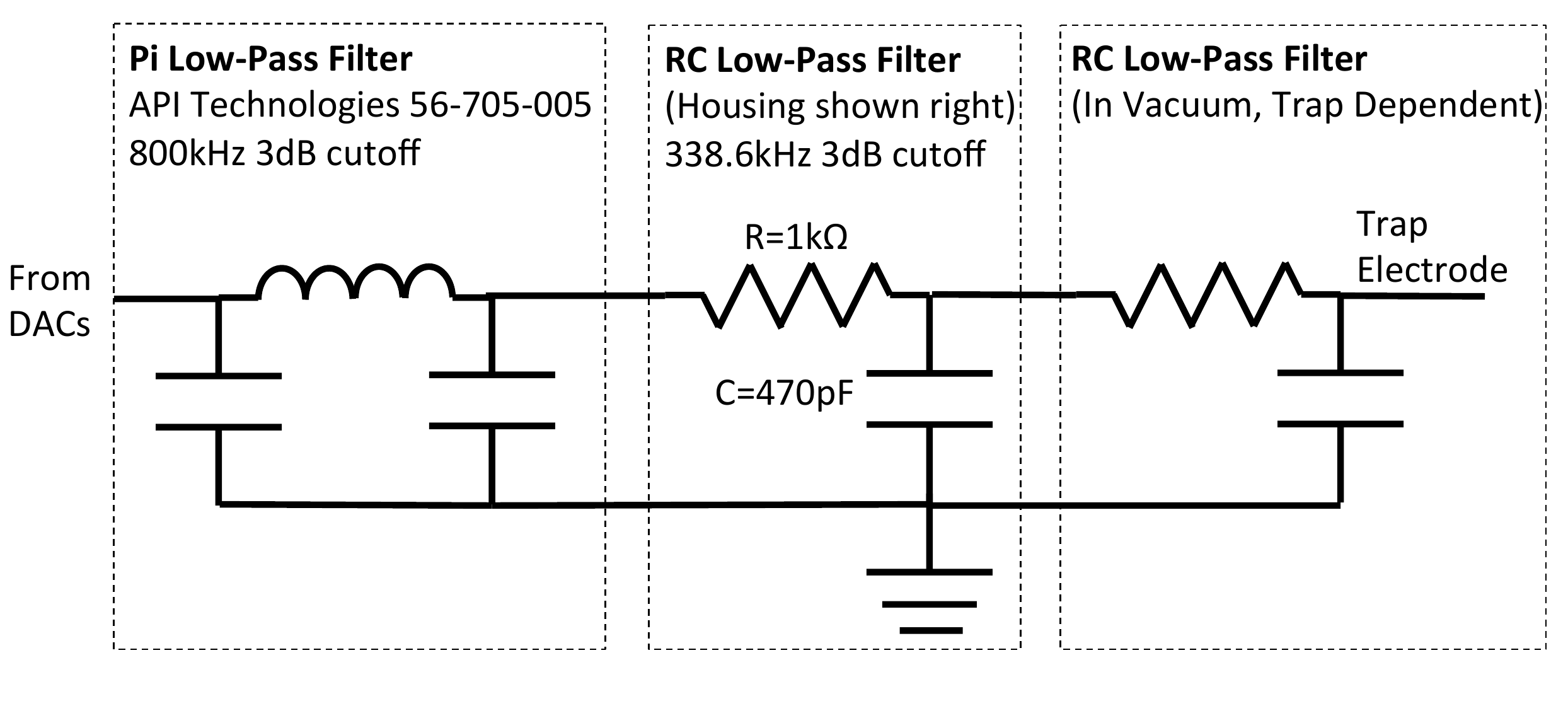}
\caption{ A schematic of the filters used between the DAC system and an ion trap. Outside of the vacuum chamber, an in-line Pi-filter is connected to an RC filter on a custom PCB that is explicitly grounded. The last set of filters are on the ion trap chip in the vacuum chamber.}
\label{fig:DACFilters}
\end{figure*}

The outputs of the DACs pass through several layers of filtering before reaching the ion trap electrodes in an effort to reduce noise near the trap frequencies. These filters reduce both noise generated by the DAC system itself and noise picked up on the wires that connect the DAC system to the ion trap. It is common practice in surface electrode ion traps to use the electrodes powered by the DACs as an AC ground for the RF voltages used inside the trap. If this RF voltage has a weak ground path, it can radiate throughout the system and generate yet more noise. Within these constraints, we found it necessary to provide a good path to ground for the system at the filters immediately outside of the electrical vacuum feedthrough.  The RC filters are housed in a Faraday cage machined out of a block of steel that is connected directly to both the trap ground and to a grounding strap in an effort to reduce the noise on the DAC outputs. The filters used have $\text{C}=470$~pF and $\text{R}=1$~k$\Omega$, with a cut off frequency $f \simeq 340$~kHz. This connector provides a low-resistance path to ground and acts to shield the lines after the filters from any additional noise. Commercial in-line filters (API Technologies 56-705-005 filters) with a cutoff frequency of 800~kHz were added to increase the filtering at high frequency and avoid any unwanted pickup near the self-resonant frequency of the capacitors used in the RC filters. Figure~\ref{fig:DACFilters} shows a circuit diagram of these filters.

\begin{figure*}
\includegraphics[width=1\columnwidth]{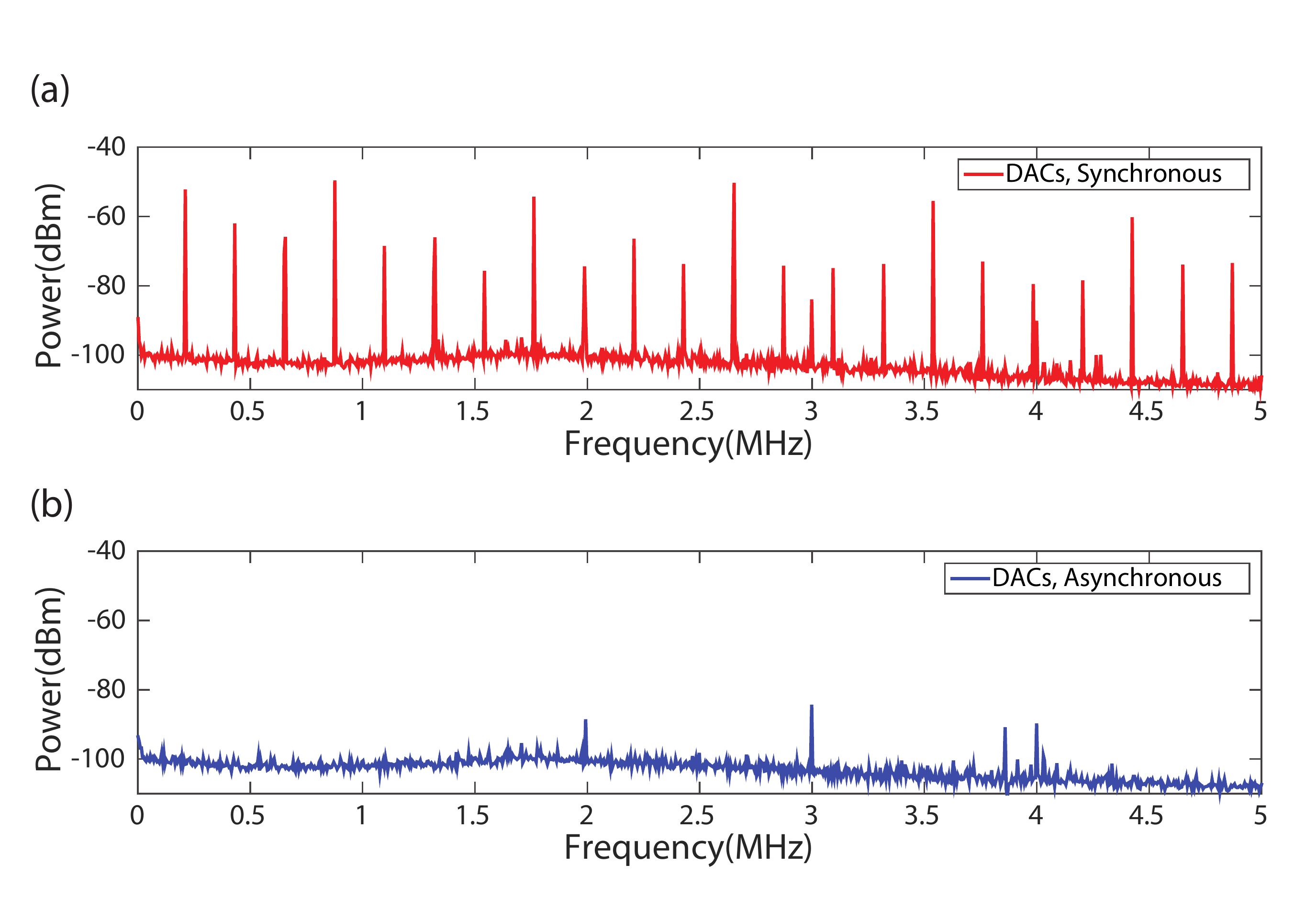}
\caption{Noise measured on the a single output channel of the DAC system at 5~V. The spectrum analyzer resolution bandwidth was 100~Hz, and the input impedance was 1~M$\Omega$. The system was run in two modes:  "synchronous mode", where the system is updated at the maximum rate of 430~kHz, and  "asynchronous mode", where the system is updated only once. (a) The synchronous mode noise spectrum is indicative of the system's performance during the fastest possible shuttling operation, while (b) the asynchronous mode noise spectrum is analogous to the system's behavior when not shuttling. These measurements do not include any filtering.}
\label{fig:DACNoise}
\end{figure*}

It is useful to evaluate the noise performance of the DAC system in two modes: synchronous mode, where the output voltages are updated at their maximum rate, and asynchronous mode, where the output voltages are updated once and left constant. In practice, the update rate can be anywhere between these two extremes, but synchronous mode is a good approximation of the behavior of the system during ion shuttling, and asynchronous mode is a good approximation for all other times. Figure \ref{fig:DACNoise} shows the noise spectrum of the output of the DAC system for both of these modes. The noise spectrum for the synchronous case clearly shows noise at the update rate (430~kHz), its half-harmonic, and its harmonics. However, much of this noise will be significantly reduced by the filters shown in figure~\ref{fig:DACFilters}.

We have demonstrated the performance of a low noise DAC system, with lower cost compared to commercially available DACs. These DACs are optimized for the common case of static ions while retaining significant shuttling capabilities.  During shuttling their performance will be slower than the DACs shown by Bowler et. al.~\cite{Bowler2012}, yet comparable to the NI PXI-6733 used by Shu et. al.~\cite{Shu2014}. Our DAC system is significantly less expensive, easier to program, and require less computational overhead than the other two alternatives.

\section{Summary and conclusion}
\label{sec:6}

As we scale up trapped ion quantum computing to a system large enough to perform significant quantum computations, the amount of classical hardware required will also grow. Here we described our efforts towards designing control hardware that is scalable and extensible while minimizing expense and spatial requirements. To this end, we have developed an offset frequency lock for CW lasers, such as those used for Doppler cooling, which allows many slave lasers to be locked to a single master laser, where each slave laser can also serve as a new master laser in a tree like structure. A single laser locked to a stable reference is sufficient to frequency stabilize a large number of slave lasers with little expense and space required for each slave laser lock. For pulsed laser used to drive Raman transitions, we have developed a pulsed laser Raman beam stabilization system that does not require access to the laser cavity, and can be implemented using simple and inexpensive low frequency components. With this lock the coherence time measured using a \Yb ion is 811 (69)~ms, which is sufficient a large scale computation containing thousands of logic gates.  Currently we are developing a general purpose PI loop system that could easily replace the pulsed laser Raman laser stabilization, intensity, and offset frequency locking controllers and software. The compact and economical system will be able to support 8 simultaneous locks with DC output, or optional off-chip DDS output. As each logic unit in the MUSIQC architecture will likely require an individual surface electrode ion trap with 100 or more DC electrodes, we have developed a low-noise, asynchronous, DAC system that compares favorably to the available commercial systems in cost, size, and flexibility.

\begin{acknowledgements}
This work was supported by the Office of the Director of National Intelligence and Intelligence Advanced Research Projects Activity through the Army Research Office.
\end{acknowledgements}

\bibliographystyle{spphys}       
\bibliography{manuscript}   

\end{document}